\documentclass[letter]{aa}
\usepackage{txfonts}
\usepackage{graphicx}
\usepackage{natbib,twoopt}
\usepackage[colorlinks,linkcolor=blue,urlcolor=blue,citecolor=black]{hyperref} %% to avoid \citeads line fills
\bibpunct{(}{)}{;}{a}{}{,}             %% natbib format for A&A and ApJ
\makeatletter
  \newcommandtwoopt{\citeads}[3][][]{\href{http://adsabs.harvard.edu/abs/#3}%
    {\def\hyper@linkstart##1##2{}%
     \let\hyper@linkend\@empty\citealp[#1][#2]{#3}}}
  \newcommandtwoopt{\citepads}[3][][]{\href{http://adsabs.harvard.edu/abs/#3}%
    {\def\hyper@linkstart##1##2{}%
     \let\hyper@linkend\@empty\citep[#1][#2]{#3}}}
  \newcommandtwoopt{\citetads}[3][][]{\href{http://adsabs.harvard.edu/abs/#3}%
    {\def\hyper@linkstart##1##2{}%
     \let\hyper@linkend\@empty\citet[#1][#2]{#3}}}
  \newcommandtwoopt{\citeyearads}[3][][]%
    {\href{http://adsabs.harvard.edu/abs/#3}
    {\def\hyper@linkstart##1##2{}%
     \let\hyper@linkend\@empty\citeyear[#1][#2]{#3}}}
  \renewcommand*\aa@pageof{, page \thepage{} of \pageref*{LastPage}} % to get rid of "Package hyperref Warning: Suppressing link with empty target on input line 92." warnings
\makeatother
\begin{document}

\title{A quantitative spectral analysis of 14 hypervelocity stars from the MMT survey} 
\author{A.~Irrgang\inst{\ref{remeis}}
        \and
        S.~Kreuzer\inst{\ref{remeis}}
        \and
        U.~Heber \inst{\ref{remeis}}
        \and 
        W.~Brown\inst{\ref{smithsonian}}
}
\institute{Dr.~Karl~Remeis-Observatory \& ECAP, Astronomical Institute, Friedrich-Alexander University Erlangen-N\"urnberg (FAU), Sternwartstr.~7, 96049 Bamberg, Germany\\ \email{andreas.irrgang@fau.de}\label{remeis}
\and
Smithsonian Astrophysical Observatory, 60 Garden Street, Cambridge, MA 02138, USA\label{smithsonian}
} 

% \date{Received  / Accepted }
\date{Received 30 April 2018 / Accepted 20 June 2018}

\abstract 
{Hypervelocity stars (HVSs) travel so fast that they may leave the Galaxy. The tidal disruption of a binary system by the supermassive black hole in the Galactic center is widely assumed to be their ejection mechanism.} 
{To test the hypothesis of an origin in the Galactic center using kinematic investigations, the current space velocities of the HVSs need to be determined. With the advent of Gaia's second data release, accurate radial velocities from spectroscopy are complemented by proper motion measurements of unprecedented quality. 
% Because the HVSs are very far away, their parallactic motion is still too small to be measured. 
Based on a new spectroscopic analysis method, we provide revised distances and stellar ages, both of which are crucial to unravel the nature of the HVSs.}
{We reanalyzed low-resolution optical spectra of 14 HVSs from the MMT HVS survey using a new grid of synthetic spectra, which account for deviations from local thermodynamic equilibrium, to derive effective temperatures, surface gravities, radial velocities, and projected rotational velocities. Stellar masses, radii, and ages were then determined by comparison with stellar evolutionary models that account for rotation. Finally, these results were combined with photometric measurements to obtain spectroscopic distances.}
{The resulting atmospheric parameters are consistent with those of main sequence stars with masses in the range 2.5 -- 5.0\,$M_\odot$. The majority of the stars rotate at fast speeds, providing further evidence for their main sequence nature. Stellar ages range from 90 to 400\,Myr and distances (with typical $1\sigma$-uncertainties of about 10--15\%) from 30 to 100\,kpc. Except for one object (B711), which we reclassify as A-type star, all stars are of spectral type B.} 
{The spectroscopic distances and stellar ages derived here are key ingredients for upcoming kinematic studies of HVSs based on Gaia proper motions.} 

          
\maketitle
\begin{figure*}
\begin{center}
\includegraphics[width=\textwidth]{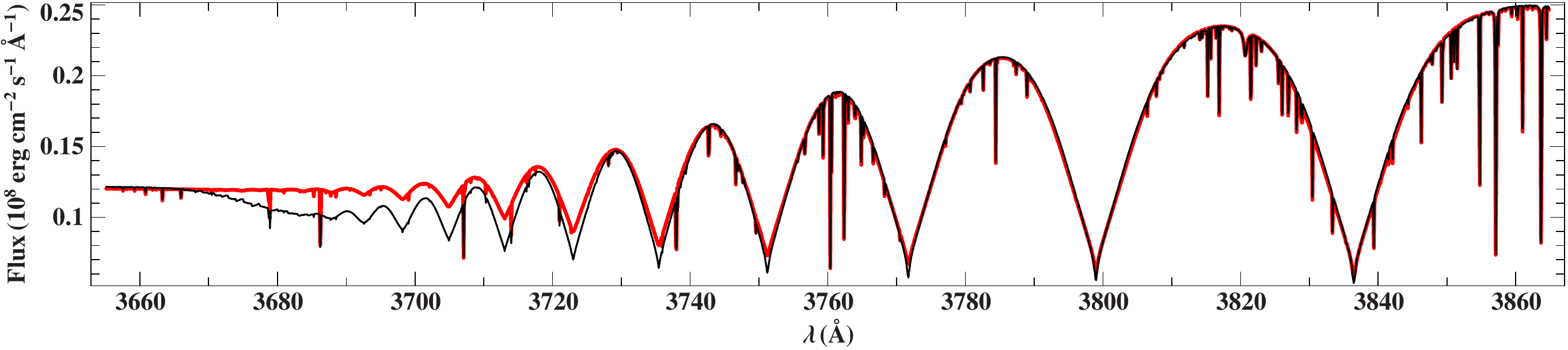}
\caption{Comparison of synthetic spectra ($T_\mathrm{eff}=12\,000$\,K, $\log(g)=4.4$, solar composition) calculated with the original {\sc Atlas12} code (thin black line) and our modified version including level dissolution (thick red line). The proper implementation of this effect allows us to exploit the region around the Balmer jump, which is an important indicator for the surface gravity, for quantitative spectral analysis, which is particularly important for the analysis of the MMT spectra given their spectral range (3600 -- 4500\,\AA).\label{fig:level_dissolution}}
\end{center}
\end{figure*}
\section{Introduction}\label{sect:intro}
Runaway stars are young early-type stars with unusual kinematics that have been found outside of star-forming regions (OB associations, open clusters). Surveys of blue stars have even revealed their presence in the Galactic halo, into which they have probably been injected from the Galactic disk by a powerful ejection mechanism. Many decades ago now, two ejection mechanisms were proposed. In the binary supernova ejection scenario \citepads{1961BAN....15..265B}, the secondary star of a close binary system is ejected when the primary explodes as a core-collapse supernova. Alternatively, dynamical interactions between (binary) stars in young, open clusters can give large kicks to the stars involved in a close encounter \citepads{1967BOTT....4...86P}.

The kinematically most extreme stars, the hypervelocity stars (HVSs), are believed to be ejected by the supermassive black hole in the Galactic center via the tidal disruption of a binary system (for a review see \citeads{2015ARA&A..53...15B}). This so-called Hills mechanism \citepads{1988Natur.331..687H} can accelerate stars to $\sim$1000\,km\,s$^{-1}$, that is, well beyond the Galactic escape velocity, making them gravitationally unbound to the Milky Way. After the first serendipitous discoveries (\citeads{2005ApJ...622L..33B}, \citeads{2005A&A...444L..61H}, \citeads{2005ApJ...634L.181E}), a systematic survey covering 12\,000 square degrees of the sky led to the discovery of almost two dozen unbound HVSs of late B type with masses between 2.5 and 4\,M$_\odot$ \citepads{2014ApJ...787...89B}. Besides the unbound HVSs, the Hills mechanism is also predicted to produce an equal number of lower velocity, i.e., probably bound, HVSs \citepads{2009ApJ...706..925B}. Indeed, \citetads{2014ApJ...787...89B} identified 16 such stars whose Galactic rest-frame velocities exceed 275\,km\,s$^{-1}$. The velocity distribution of young stars in the Galactic halo is therefore expected to be produced by an overlap of the three different acceleration mechanisms listed above, at least for the lower-velocity tail of HVSs ($<700$\,km\,s$^{-1}$) where extreme runaway stars start to contribute (\citeads{2008A&A...483L..21H}, \citeads{2008A&A...480L..37P}, \citeads{2010ApJ...711..138I}, \citeads{2011MNRAS.411.2596S}). In order to distinguish between the scenarios, kinematic studies have to be used \citepads{2014ApJ...793..122K}. Proper motions measured with the Hubble Space Telescope (HST) turned out to be not accurate enough to calculate trajectories with sufficiently small uncertainties to pinpoint the origin in the Galactic plane \citepads{2015ApJ...804...49B}. With its second data release, Gaia now provides more accurate proper motions. 
% However, even end-of-mission parallaxes will not be sufficiently accurate at very large distances (up to 100\,kpc) typical of HVSs. Spectroscopic distances remain the only option even in the Gaia era. 
Therefore, it is of utmost importance to determine spectroscopic distances that    are as accurate and precise as possible in order to exploit the excellent quality of Gaia's proper motions.

The MMT HVS survey by \citetads{2014ApJ...787...89B} discovered 38 late B-type HVSs, 16 of which are considered to be bound to the Galaxy. We selected the 14 most interesting ones, that is, those stars which have proper motions measured by HST. The corresponding MMT spectra cover the very blue optical range from 3600 to 4500\,\AA, which nicely traces the confluence of the Balmer series. However, the modeling of this region is challenging because a proper treatment of the dissolution of the hydrogen levels is required. We used a new generation of model atmospheres and synthetic spectra which account for this effect (Sect.~\ref{sect:atmos}) to reanalyze the MMT spectra of \citetads{2014ApJ...787...89B},  allowing us to derive effective temperatures and surface gravities with significantly improved accuracy (Sect.~\ref{sect:analysis}). Additionally, radial and projected rotational velocities are determined during our simultaneous fitting procedure. The latter are of particular importance in order to distinguish between blue horizontal-branch (BHB) and B-type main sequence stars, which populate the same region in the Hertzsprung-Russell-diagram and are, thus, indistinguishable by means of effective temperature and surface gravity alone \citepads{2008ASPC..392..167H}. High projected rotational velocities are indicative of a main sequence nature. By comparison with evolutionary tracks, we derived stellar masses and ages (Sect.~\ref{sect:mass_age}) and, in combination with photometry, spectroscopic distances (Sect.~\ref{sect:photo_analysis}). We conclude with a brief glimpse into the Gaia era (Sect.\ref{sect:outlook}).
\section{Model atmospheres and synthetic spectra}\label{sect:atmos}
Synthetic spectra for B-type stars are calculated from the structure of a line-blanketed, plane-parallel, homogeneous, and hydrostatic model atmosphere in local thermodynamic equilibrium (LTE) computed with the {\sc Atlas12} code \citepads{1996ASPC..108..160K}. Departures from LTE are allowed for by calculating atomic population numbers using the {\sc Detail} code \citepads{1981PhDT.......113G} to solve the coupled radiative transfer and statistical equilibrium equations. The synthetic spectrum is then computed with {\sc Surface} \citepads{1981PhDT.......113G} making use of the departure coefficients and the best line-broadening data available. This so-called {\sc Atlas}, {\sc Detail}, {\sc Surface} (ADS) or hybrid LTE/non-LTE approach is described in detail by \citetads{2011JPhCS.328a2015P}. To model spectra of A-type stars and cooler, we make use of the pure LTE spectrum synthesis code {\sc Synthe} \citepads{1993sssp.book.....K}.

Recently, the {\sc Atlas12}, {\sc Detail}, {\sc Surface}, and {\sc Synthe} codes have been improved by one of us (A.~Irrgang). The occupation probability formalism \citepads{1988ApJ...331..794H} for hydrogen and ionized helium --~following the description given by \citetads{1994A&A...282..151H}~-- as well as state-of-the-art line broadening tables for hydrogen \citepads{2009ApJ...696.1755T} have been implemented. Both are of particular importance to model the Balmer jump as demonstrated in Fig.~\ref{fig:level_dissolution}. Moreover, non-LTE effects on the atmospheric structure are now considered in the sense that departure coefficients for hydrogen and helium are passed back from {\sc Detail} to {\sc Atlas12} to refine the atmospheric structure iteratively. Among other things, this ensures a more realistic representation of the spectral energy distribution (SED).

A grid of synthetic spectra with solar chemical composition based on the ADS approach has been computed for effective temperatures $T_\mathrm{eff}$ ranging from 9000 to 16\,000\,K in steps of 250\,K and surface gravities $\log(g)$ between 3.0 and 4.6 in steps of 0.2. To account for cooler stars, a similar grid based on {\sc Atlas12} and {\sc Synthe} has been computed covering effective temperatures between 7000 and 10\,000\,K in steps of 200\,K. 
\section{Atmospheric parameters and projected rotational velocities}\label{sect:analysis}
\begin{figure*}
\begin{center}
\includegraphics[width=1\textwidth]{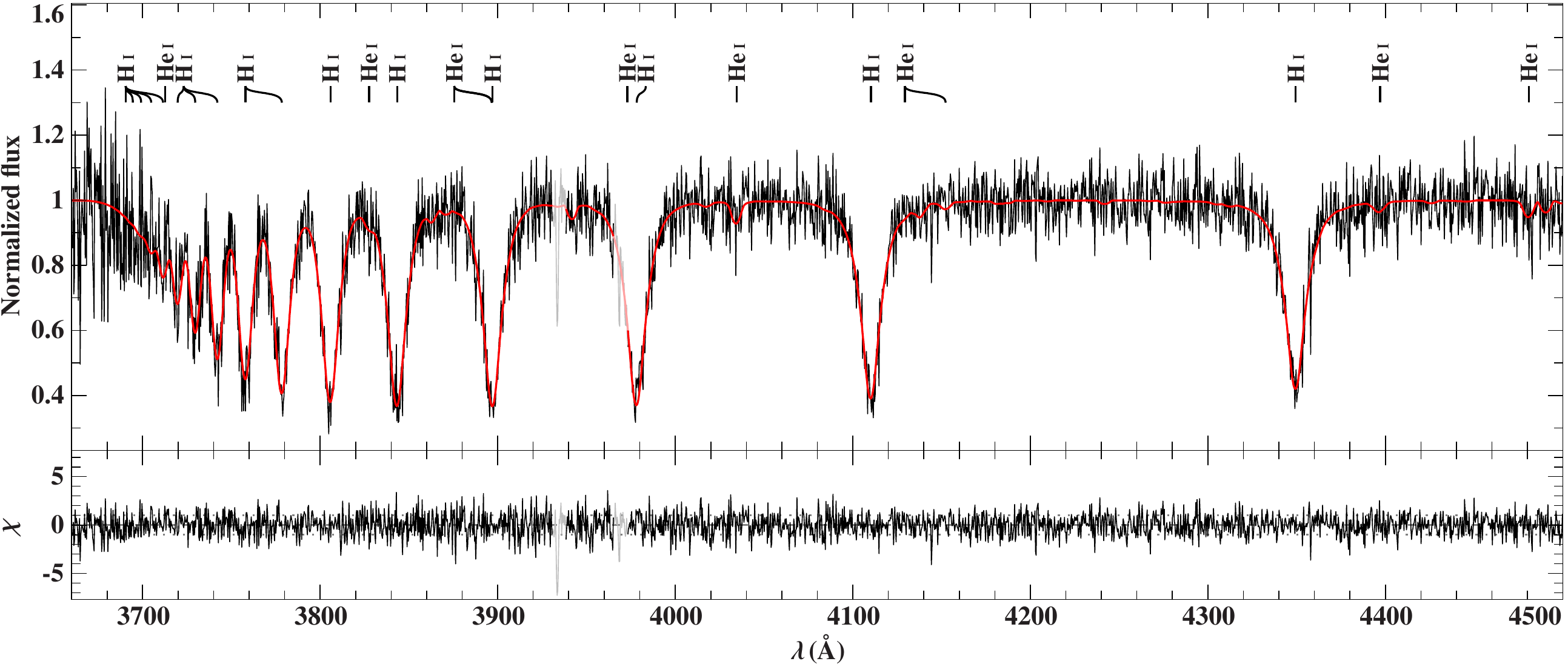}
\caption{Comparison of a normalized observed MMT spectrum (black) with the best-fitting synthetic spectrum (red) in the case of HVS\,9. Lines of hydrogen and helium are labeled for reference. Regions contaminated by interstellar Ca\,{\sc ii} lines were excluded and appear in light colors. Residuals $\chi$ are shown in the lower panel.\label{fig:hvs9}}
\end{center}
\end{figure*}
\begin{figure}
\begin{center}
\includegraphics[width=0.49\textwidth]{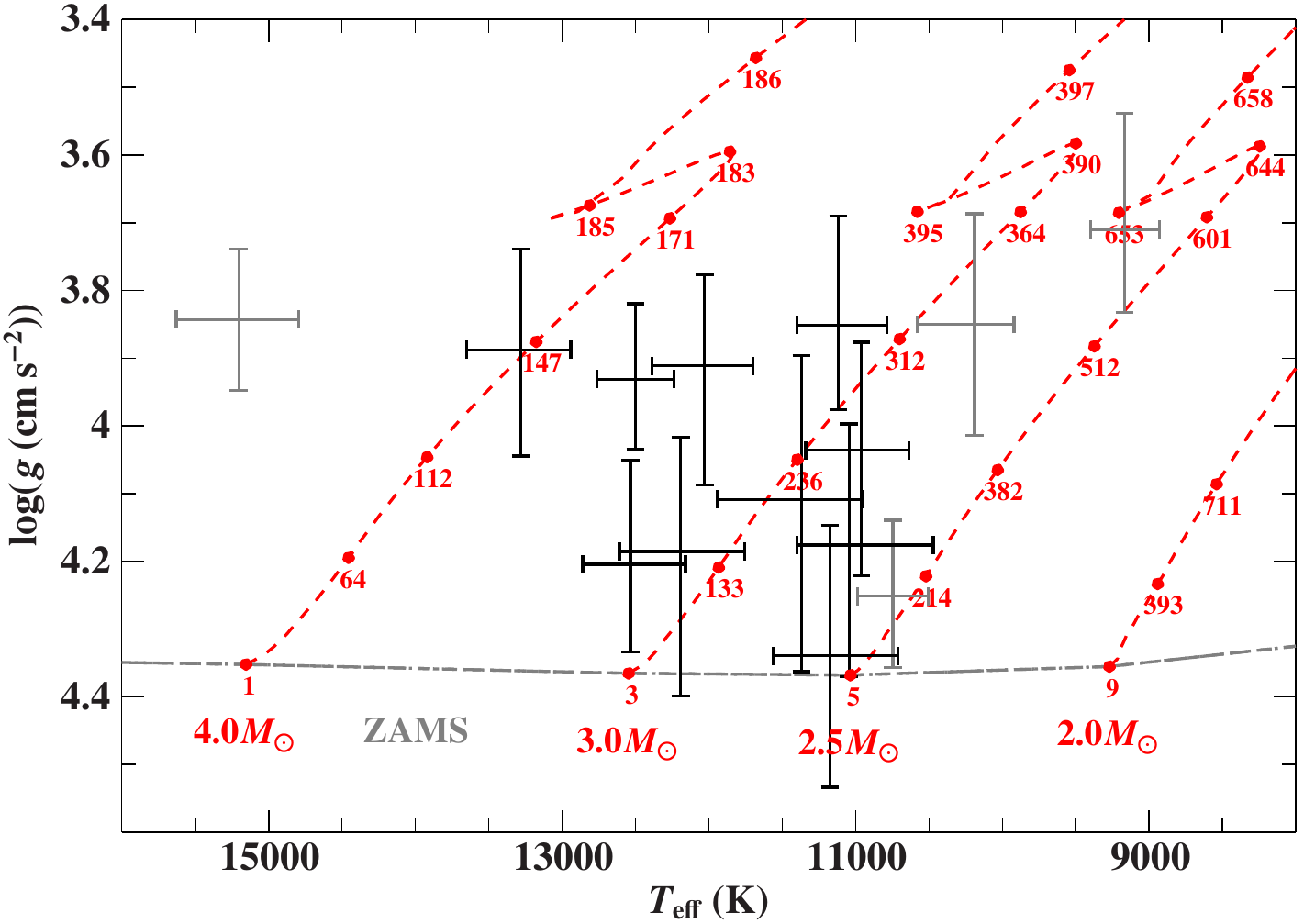}
\caption{Position of the 14 program stars in a ($T_\mathrm{eff}$, $\log(g)$) diagram (stars with HVS identifier are black, others are gray). Evolutionary tracks for rotating ($\Omega/\Omega_\mathrm{crit} = 0.4$) main sequence stars of solar metallicity and different initial masses \citepads{2013A&A...553A..24G} are overlaid in red. Red filled circles and numbers mark the age in Myr. The locus of the zero-age main sequence (ZAMS) is indicated as a gray dashed line. Error bars indicate 99\%-confidence limits and cover statistical and systematic uncertainties.\label{fig:kiel_diagram}}
\end{center}
\end{figure}
The quantitative analysis strategy as described by \citetads{2014A&A...565A..63I} was applied to simultaneously fit the entire spectral range to derive $T_\mathrm{eff}$, $\log(g)$, the radial velocity $\varv_{\mathrm{rad}}$, and the projected rotational velocity $\varv\sin(i)$. An exemplary fit to one of the MMT spectra of HVS\,9 is shown in Fig.~\ref{fig:hvs9}. The results are listed in Table~\ref{table:parameters} and visualized in Fig.~\ref{fig:kiel_diagram}. While for most of the stars in the sample, the effective temperatures agree quite well with the previous results by \citetads{2015ApJ...804...49B}, there are three objects (HVS\,4, HVS\,12, B485) that show significant differences of $\sim 1000$\,K and one (B711) of $\sim 2000$\,K. Given the wealth of spectral lines in its spectrum, the program star B711 turned out to be an early A-type instead of a late B-type star. The new surface gravities do not agree well with previous results. However, this is expected because the more sophisticated treatment of level dissolution in our new models alters the profiles as well as the confluence of the Balmer lines. Our surface gravities scatter less and nicely match the range predicted by stellar evolution models for the main sequence, resolving previous issues with stars being too close or even below the ZAMS. Radial velocities are consistent with those published by \citet{2015ApJ...804...49B} except for B485, for which our result ($422.9^{+4.5}_{-3.4}$\,km\,s$^{-1}$) is slightly higher ($408.1\pm 4.8$\,km\,s$^{-1}$). The projected rotational velocities exceed 100\,km\,s$^{-1}$ for all but five stars. On average, our results are somewhat lower than those published by \citetads{2015ApJ...804...49B} except for HVS\,4 ($138^{+34}_{-40}$\,km\,s$^{-1}$ vs.\ $77\pm40$\,km\,s$^{-1}$). The fast rotation indicates that those stars are likely main sequence stars. The projected rotational velocity of HVS\,10 and HVS\,12 is consistent with zero, which could be explained by a low inclination of the rotational axis.

Three stars of the sample have
previously been studied with high-resolution spectra. \citetads{2008A&A...488L..51P} used the ADS approach to analyze HVS\,7. Their parameters ($T_\mathrm{eff}=12\,000 \pm 500$\,K, $\log(g)=3.8\pm0.1$, $\varv\sin(i)=55 \pm 2$\,km\,s$^{-1}$) agree very well with our values presented in Table~\ref{table:parameters}. Similarly, the parameters derived by \citetads{2008ApJ...685L..47L} for HVS\,8 ($T_\mathrm{eff}=11\,000 \pm 1000$\,K, $\log(g)=3.75\pm0.25$, $\varv\sin(i)=260 \pm 70$\,km\,s$^{-1}$) and by \citetads{2012ApJ...754L...2B} for HVS\,5 ($T_\mathrm{eff}=12\,000 \pm 350$\,K, $\log(g)=3.89\pm0.13$, $\varv\sin(i)=133 \pm 7$\,km\,s$^{-1}$) are consistent with ours. This good agreement gives us confidence that the spectral coverage and resolution of the MMT spectra are sufficient to derive accurate parameters.
\section{Evolutionary masses and stellar ages}\label{sect:mass_age}
Figure~\ref{fig:kiel_diagram} shows the comparison between derived atmospheric parameters and predictions of rotating stellar models of solar metallicity for the main sequence \citepads{2013A&A...553A..24G}. The resulting atmospheric parameters are consistent with the stars being main sequence stars with masses between 2.5 and 5.0\,M$_\odot$. Stellar masses $M$, ages $\tau$, radii $R$, luminosities $L$, and ratios of actual angular velocity to critical velocity $\Omega/\Omega_\mathrm{crit}$ are derived by interpolation of the evolutionary tracks and are listed in Table~\ref{table:parameters}. Lacking other indicators, we use the projected rotational velocity (multiplied by a statistical factor of $4/\pi$ based on the simple assumption of isotropic $\sin(i)$ values) as proxy for the equatorial rotation and, hence, for stellar rotation. 
\section{Spectroscopic distances and interstellar reddening}\label{sect:photo_analysis}
\begin{figure}
\begin{center}
\includegraphics[width=0.49\textwidth]{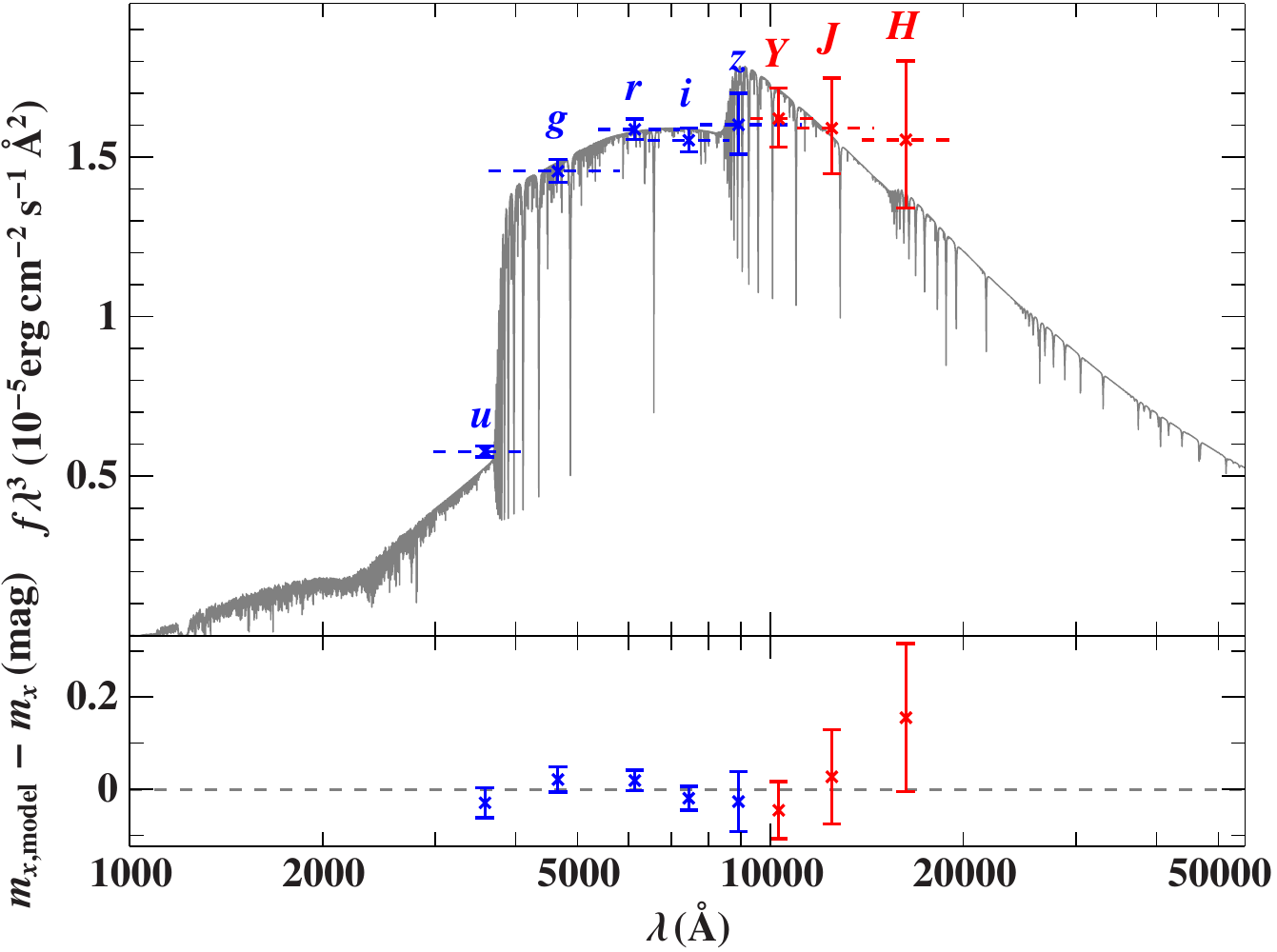}
\caption{Comparison of synthetic and observed photometry for HVS\,9: The \textit{top panel} shows the SED. The colored data points (SDSS-DR14: blue; UKIDDS-DR9: red) are filter-averaged fluxes which were converted from observed magnitudes (the respective filter widths are indicated by the dashed horizontal lines), while the gray solid line represents a model that is based on the spectroscopic parameters given in Table~\ref{table:parameters}. The flux is multiplied with the wavelength to the power of three to reduce the steep slope of the SED on such a wide wavelength range. The residual panel at the \textit{bottom} shows the differences between synthetic and observed magnitudes. We note that only the angular diameter and the color excess were fitted.\label{fig:sed_fit}}
\end{center}
\end{figure}
Photometric measurements allow additional parameters to be derived, namely the stellar angular diameter and the interstellar reddening. The angular diameter, in turn, grants access to the distance because the stellar radius is already known (see Sect.~\ref{sect:mass_age}). We constructed SEDs from optical (SDSS-DR14: \citeads{2017AJ....154...28B}) and infrared (UKIDSS-DR9: \citeads{2007MNRAS.379.1599L}; ALLWISE: \citeads{2014yCat.2328....0C}) photometry. Synthetic SEDs (based on the atmospheric parameters derived via spectroscopy; see Table~\ref{table:parameters}) were then calculated with our improved version of {\sc Atlas12} and fitted to the observed ones by adjusting the angular diameter and the color excess $E(B-V)$ (for details see \citeads{2018OAst...27...35H}). The extinction curve by  \citetads{1999PASP..111...63F} was used to compute the wavelength-dependent reddening factor $10^{-0.4 A(\lambda)}$, with which the flux has to be multiplied, from the fitting parameter $E(B-V)$ and the fixed extinction parameter $R_V = A(V) / E(B-V) = 3.1$. As an example, Fig.~\ref{fig:sed_fit} shows the result of such an exercise in the case of HVS\,9. The resulting values for the color excess and stellar distance are listed in Table \ref{table:parameters}. The error budget is dominated by the uncertainties of the atmospheric parameters, in particular by $\log(g)$. The resulting distances range from 10 to 100\,kpc and have typical $1\sigma$-uncertainties of about 10--15\%. 
% For comparison, we include parallaxes from Gaia's second data release \citepads{2018arXiv180409365G} in Table~\ref{table:parameters}. For all but one star, the uncertainties are larger than their respective values and/or the values are negative, which shows that no parallax is significant except maybe the one for B733. However, we strongly doubt that this measurement is reliable because it implies a surface gravity that would place B733 below the zero-age main sequence. This would require the object to be a BHB star, which is at odds with its high projected rotational velocity.
%
\section{The Gaia era}\label{sect:outlook}
Gaia astrometry will have an enormous impact on stellar astrophysics and Galactic astronomy. The mission will provide proper motions of unprecedented accuracy. For instance, a comparison between HVS proper motions measured with HST and those measured with Gaia is given by \citetads{2018MNRAS.476.4697M} for predicted end-of-mission performance and by \citetads{2018arXiv180504184B} for the second data release. For the first time, it will be possible to derive significant constraints on the place of origin as well as on the ejection mechanism of HVSs. 
% However, because the HVS stars are so faint ($\sim$18--20\,mag) and far away (30--100\,kpc), Gaia parallaxes will not be meaningful. 
In order to derive precise transversal velocities, we aimed at determining spectroscopic distances for 14 HVSs by reanalyzing their spectra from the MMT HVS survey using models with improved spectral synthesis of the Balmer line series. Besides revised atmospheric parameters and radial and projected rotational velocities, we provide more realistic stellar masses and ages. The resulting atmospheric distances (accurate to 10--15\% at $1\sigma$ confidence) will be most useful to obtain the three-dimensional (3D) space motions of the stars by combining them with proper motions from Gaia DR2. The accuracy of the spectroscopic distances are mostly limited by systematic uncertainties in the determination of the surface gravity. Gaia parallaxes of nearby B-type reference stars will allow surface gravities to  be cross-checked and, thus, help to reduce the systematic uncertainties. High-resolution spectroscopy is required to study the chemical composition of the stars. Last but not least, our derived stellar ages can be compared to kinematic times of flight to further constrain the origin of the HVSs. 
\begin{sidewaystable*}
\begin{center}
\footnotesize
\setlength{\tabcolsep}{0.109cm}
\renewcommand{\arraystretch}{1.5}
\caption{\label{table:parameters} Atmospheric and stellar parameters of the program stars.}
\begin{tabular}{lrrrrrrrrrrrrrrrrrrrrrrrrrrrrrrrrrrr}
\hline\hline
Object & & \multicolumn{2}{c}{$T_{\mathrm{eff}}$} & & \multicolumn{2}{c}{$\log(g)$} & & \multicolumn{2}{c}{$\varv_{\mathrm{rad}}$} & & \multicolumn{2}{c}{$\varv\sin(i)$} & & \multicolumn{2}{c}{$M$} & & \multicolumn{2}{c}{$\tau$} & & \multicolumn{2}{c}{$\log(L/L_\odot)$} & & \multicolumn{2}{c}{$R$} & & \multicolumn{2}{c}{$\Omega/\Omega_\mathrm{crit}$}  & & \multicolumn{2}{c}{$E(B-V)$} & & \multicolumn{2}{c}{$d$} & $1\sigma$ \\
\cline{3-4} \cline{6-7} \cline{9-13} \cline{15-16} \cline{18-19} \cline{21-22} \cline{24-25} \cline{27-28} \cline{30-31} \cline{33-35} 
& & \multicolumn{2}{c}{(K)} & & \multicolumn{2}{c}{(cgs)} & & \multicolumn{5}{c}{$(\mathrm{km\,s^{-1}})$} & & \multicolumn{2}{c}{($M_\odot$)} & & \multicolumn{2}{c}{(Myr)} & & & & &  \multicolumn{2}{c}{($R_\odot$)} & & & & & \multicolumn{2}{c}{(mag)} & & \multicolumn{3}{c}{(kpc)} \\
\hline
HVS\,1 (\object{SDSS\,J090744.99$+$024506.9})  &         & $11\,120$ & $^{+280}_{-340}$ &         & $3.85$ & $^{+0.13}_{-0.16}$ &         & $829.7$ &             $^{+5.5}_{-5.7}$ &         & $160$ &             $^{+37}_{-23}$ &         & $3.2$ & $^{+0.3}_{-0.3}$ &         & $272$ &             $^{+29}_{-28}$ &         & $2.22$ & $^{+0.20}_{-0.15}$ &         & $3.5$ & $^{+1.1}_{-0.6}$ &         & $0.75$ & $^{+0.05}_{-0.05}$ &         & $0.099$ & $^{+0.087}_{-0.069}$ &          & $99.3$ &           $^{+39.2}_{-23.6}$ & $^{+15.2}_{-\phantom{0}9.2}$  \\ % &          &  $0.2396$ & $\pm0.8591$ \\
HVS\,4 (\object{SDSS\,J091301.01$+$305119.8})  &         & $13\,280$ & $^{+370}_{-340}$ &         & $3.89$ & $^{+0.16}_{-0.16}$ &         & $604.6$ &             $^{+7.1}_{-6.6}$ &         & $138$ &             $^{+34}_{-40}$ &         & $4.0$ & $^{+0.4}_{-0.4}$ &         & $150$ &             $^{+14}_{-26}$ &         & $2.59$ & $^{+0.21}_{-0.22}$ &         & $3.8$ & $^{+1.0}_{-0.9}$ &         & $0.64$ & $^{+0.03}_{-0.05}$ &         & $0.005$ & $^{+0.016}_{-0.005}$ &          & $78.3$ &           $^{+22.2}_{-18.5}$ &             $^{+8.6}_{-7.2}$  \\ % &          &  $0.3632$ & $\pm0.5792$ \\
HVS\,5 (\object{SDSS\,J091759.48$+$672238.3})  &         & $12\,530$ & $^{+330}_{-380}$ &         & $4.20$ & $^{+0.14}_{-0.15}$ &         & $542.5$ &             $^{+7.4}_{-7.6}$ &         & $131$ &             $^{+30}_{-33}$ &         & $3.3$ & $^{+0.2}_{-0.3}$ &         &  $97$ &             $^{+78}_{-95}$ &         & $2.09$ & $^{+0.20}_{-0.18}$ &         & $2.4$ & $^{+0.6}_{-0.5}$ &         & $0.61$ & $^{+0.04}_{-0.18}$ &         & $0.109$ & $^{+0.033}_{-0.036}$ &          & $31.2$ &             $^{+8.3}_{-6.4}$ &             $^{+3.2}_{-2.5}$  \\ % &          & $-0.0756$ & $\pm0.1861$ \\
HVS\,6 (\object{SDSS\,J110557.45$+$093439.5}   &         & $12\,190$ & $^{+420}_{-440}$ &         & $4.19$ & $^{+0.21}_{-0.18}$ &         & $619.3$ & $^{+\phantom{0}9.7}_{-10.5}$ &         & $125$ &             $^{+44}_{-89}$ &         & $3.1$ & $^{+0.2}_{-0.4}$ &         & $142$ & $^{+\phantom{0}85}_{-139}$ &         & $2.04$ & $^{+0.17}_{-0.27}$ &         & $2.3$ & $^{+0.7}_{-0.6}$ &         & $0.59$ & $^{+0.07}_{-0.59}$ &         & $0.033$ & $^{+0.065}_{-0.033}$ &          & $57.7$ &           $^{+17.1}_{-18.5}$ &             $^{+6.6}_{-7.2}$  \\ % &          & $-0.7858$ & $\pm0.3950$ \\
HVS\,7 (\object{SDSS\,J113312.12$+$010824.9})  &         & $12\,500$ & $^{+270}_{-270}$ &         & $3.93$ & $^{+0.11}_{-0.12}$ &         & $524.0$ &             $^{+4.2}_{-3.8}$ &         &  $58$ &             $^{+25}_{-29}$ &         & $3.5$ & $^{+0.4}_{-0.3}$ &         & $185$ &             $^{+17}_{-25}$ &         & $2.39$ & $^{+0.18}_{-0.18}$ &         & $3.4$ & $^{+0.7}_{-0.6}$ &         & $0.31$ & $^{+0.07}_{-0.07}$ &         & $0.006$ & $^{+0.021}_{-0.006}$ &          & $48.2$ & $^{+11.1}_{-\phantom{0}9.5}$ &             $^{+4.3}_{-3.7}$  \\ % &          & $-0.1803$ & $\pm0.2006$ \\
HVS\,8 (\object{SDSS\,J094214.04$+$200322.1})  &         & $10\,960$ & $^{+380}_{-330}$ &         & $4.04$ & $^{+0.19}_{-0.17}$ &         & $499.6$ &             $^{+8.8}_{-8.6}$ &         & $282$ &             $^{+26}_{-69}$ &         & $2.9$ & $^{+0.3}_{-0.2}$ &         & $226$ & $^{+\phantom{0}60}_{-131}$ &         & $1.98$ & $^{+0.18}_{-0.21}$ &         & $2.7$ & $^{+0.8}_{-0.6}$ &         & $0.95$ & $^{+0.01}_{-0.01}$ &         & $0.026$ & $^{+0.039}_{-0.026}$ &          & $37.2$ & $^{+11.4}_{-\phantom{0}9.2}$ &             $^{+4.4}_{-3.6}$  \\ % &          & $-0.2054$ & $\pm0.2353$ \\
HVS\,9 (\object{SDSS\,J102137.08$-$005234.8})  &         & $12\,030$ & $^{+360}_{-340}$ &         & $3.91$ & $^{+0.18}_{-0.14}$ &         & $622.0$ &             $^{+7.8}_{-7.6}$ &         & $261$ &             $^{+35}_{-28}$ &         & $3.5$ & $^{+0.3}_{-0.3}$ &         & $175$ &             $^{+21}_{-60}$ &         & $2.35$ & $^{+0.15}_{-0.24}$ &         & $3.5$ & $^{+0.7}_{-0.9}$ &         & $0.95$ & $^{+0.01}_{-0.01}$ &         & $0.106$ & $^{+0.049}_{-0.036}$ &          & $66.6$ &           $^{+16.0}_{-17.8}$ &             $^{+6.2}_{-7.0}$  \\ % &          &  $0.1509$ & $\pm0.4459$ \\
HVS\,10 (\object{SDSS\,J120337.85$+$180250.4}) &         & $11\,040$ & $^{+370}_{-570}$ &         & $4.18$ & $^{+0.20}_{-0.19}$ &         & $462.0$ & $^{+12.9}_{-\phantom{0}5.6}$ &         &  $55$ &             $^{+45}_{-59}$ &         & $2.7$ & $^{+0.2}_{-0.3}$ &         & $210$ &           $^{+127}_{-206}$ &         & $1.82$ & $^{+0.20}_{-0.25}$ &         & $2.2$ & $^{+0.8}_{-0.6}$ &         & $0.29$ & $^{+0.09}_{-0.22}$ &         & $0.047$ & $^{+0.050}_{-0.047}$ &          & $54.2$ &           $^{+16.0}_{-13.8}$ &             $^{+6.2}_{-5.4}$  \\ % &          & $-1.0550$ & $\pm0.4744$ \\
HVS\,12 (\object{SDSS\,J105009.60$+$031550.7}) &         & $11\,170$ & $^{+400}_{-460}$ &         & $4.34$ & $^{+0.20}_{-0.20}$ &         & $545.0$ &             $^{+8.7}_{-8.6}$ &         &   $0$ &   $^{+46}_{-\phantom{0}0}$ &         & $2.5$ & $^{+0.2}_{-0.2}$ &         &  $90$ & $^{+197}_{-\phantom{0}86}$ &         & $1.64$ & $^{+0.22}_{-0.14}$ &         & $1.8$ & $^{+0.6}_{-0.3}$ &         & $0.00$ & $^{+0.01}_{-0.00}$ &         & $0.076$ & $^{+0.095}_{-0.076}$ &          & $51.7$ &           $^{+23.1}_{-15.8}$ &             $^{+9.0}_{-6.1}$  \\ % &          & $-0.1603$ & $\pm0.6615$ \\
HVS\,13 (\object{SDSS\,J105248.31$-$000133.9}) &         & $11\,370$ & $^{+580}_{-420}$ &         & $4.11$ & $^{+0.26}_{-0.22}$ &         & $568.1$ &           $^{+13.1}_{-14.2}$ &         & $166$ & $^{+\phantom{0}34}_{-105}$ &         & $2.9$ & $^{+0.4}_{-0.3}$ &         & $200$ & $^{+\phantom{0}83}_{-197}$ &         & $1.97$ & $^{+0.28}_{-0.29}$ &         & $2.5$ & $^{+1.1}_{-0.8}$ &         & $0.76$ & $^{+0.01}_{-0.26}$ &         & $0.037$ & $^{+0.070}_{-0.037}$ &          & $95.2$ &           $^{+45.4}_{-32.5}$ &           $^{+17.7}_{-12.7}$  \\ % &          & $-0.1322$ & $\pm1.2538$ \\
B434 (\object{SDSS\,J110224.37$+$025002.8})    &         & $10\,190$ & $^{+390}_{-270}$ &         & $3.85$ & $^{+0.17}_{-0.17}$ &         & $445.5$ &             $^{+6.3}_{-5.8}$ &         & $101$ &             $^{+35}_{-24}$ &         & $2.8$ & $^{+0.2}_{-0.2}$ &         & $402$ &             $^{+41}_{-58}$ &         & $2.01$ & $^{+0.18}_{-0.17}$ &         & $3.3$ & $^{+0.9}_{-0.8}$ &         & $0.54$ & $^{+0.08}_{-0.12}$ &         & $0.072$ & $^{+0.053}_{-0.035}$ &          & $40.5$ & $^{+12.1}_{-\phantom{0}9.5}$ &             $^{+4.7}_{-3.7}$  \\ % &          &  $0.0383$ & $\pm0.2139$ \\
B485 (\object{SDSS\,J101018.82$+$302028.1})    &         & $15\,200$ & $^{+370}_{-410}$ &         & $3.89$ & $^{+0.11}_{-0.13}$ &         & $422.9$ &             $^{+4.5}_{-3.4}$ &         &  $69$ &             $^{+27}_{-19}$ &         & $4.8$ & $^{+0.6}_{-0.4}$ &         &  $94$ &             $^{+12}_{-11}$ &         & $2.91$ & $^{+0.23}_{-0.15}$ &         & $4.1$ & $^{+1.2}_{-0.5}$ &         & $0.35$ & $^{+0.07}_{-0.05}$ &         & $0.018$ & $^{+0.010}_{-0.018}$ &          & $33.3$ &             $^{+9.4}_{-4.3}$ &             $^{+3.7}_{-1.7}$  \\ % &          & $-0.0884$ & $\pm0.0892$ \\
B711 (\object{SDSS\,J142001.94$+$124404.8})    &         &    $9170$ & $^{+230}_{-250}$ &         & $3.71$ & $^{+0.13}_{-0.18}$ &         & $271.0$ &             $^{+2.9}_{-3.5}$ &         &   $0$ &   $^{+39}_{-\phantom{0}0}$ &         & $2.8$ & $^{+0.1}_{-0.3}$ &         & $393$ & $^{+151}_{-\phantom{0}42}$ &         & $1.97$ & $^{+0.17}_{-0.14}$ &         & $3.9$ & $^{+1.0}_{-0.7}$ &         & $0.00$ & $^{+0.39}_{-0.00}$ &         & $0.009$ & $^{+0.026}_{-0.009}$ &          & $28.5$ &             $^{+8.0}_{-5.5}$ &             $^{+3.1}_{-2.2}$  \\ % &          & $-0.0788$ & $\pm0.1048$ \\
B733 (\object{SDSS\,J144955.58$+$310351.4})    &         & $10\,750$ & $^{+240}_{-250}$ &         & $4.23$ & $^{+0.12}_{-0.11}$ &         & $350.8$ &             $^{+4.1}_{-3.5}$ &         & $223$ &             $^{+14}_{-25}$ &         & $2.6$ & $^{+0.2}_{-0.3}$ &         & $123$ &           $^{+102}_{-119}$ &         & $1.71$ & $^{+0.11}_{-0.18}$ &         & $2.1$ & $^{+0.3}_{-0.5}$ &         & $0.95$ & $^{+0.01}_{-0.39}$ &         & $0.024$ & $^{+0.036}_{-0.024}$ &          &  $9.9$ &             $^{+1.7}_{-2.1}$ &             $^{+0.7}_{-0.9}$  \\ % &          &  $0.1857$ & $\pm0.0555$ \\
\hline
\end{tabular}
\tablefoot{Uncertainties in the atmospheric parameters are the quadratic sums of statistical (99\%-confidence limits) and systematic uncertainties. Systematic uncertainties cover only the effects induced by additional variations of $2\%$ in $T_{\mathrm{eff}}$ and $0.1$ in $\log(g)$ and are formally taken to be 99\%-confidence limits (see \citeads{2014A&A...565A..63I} for details). Uncertainties in the stellar parameters cover the effects induced by varying $T_{\mathrm{eff}}$ and $\log(g)$ in the given uncertainty ranges and, therefore, are also formally taken to be 99\%-confidence limits. For comparison, distance uncertainties are also given as $1\sigma$ uncertainties, i.e., as 68\%-confidence limits, by dividing the 99\%-confidence limits by a factor of $2.576$. Photometric uncertainties are included in the error budget for the distance. 
%For reference, parallaxes $\Pi$ from the second data release of Gaia are given as well.
The atmospheric parameters of B711 are based on {\sc Atlas12}/{\sc Synthe}.
}
\end{center}
\end{sidewaystable*}
\begin{acknowledgements}
We thank John E.\ Davis for the development of the {\sc slxfig} module used to prepare the figures in this paper. % SLANG XFIG
This work is based in part on data obtained as part of the UKIRT Infrared Deep Sky Survey. % UKIDSS
This publication makes use of data products from the Wide-field Infrared Survey Explorer, which is a joint project of the University of California, Los Angeles, and the Jet Propulsion Laboratory/California Institute of Technology, funded by the National Aeronautics and Space Administration. % ALLWISE
Funding for the Sloan Digital Sky Survey (SDSS) has been provided by the Alfred P. Sloan Foundation, the Participating Institutions, the National Aeronautics and Space Administration, the National Science Foundation, the U.S. Department of Energy, the Japanese Monbukagakusho, and the Max Planck Society. The SDSS Web site is http://www.sdss.org/. The SDSS is managed by the Astrophysical Research Consortium (ARC) for the Participating Institutions. The Participating Institutions are The University of Chicago, Fermilab, the Institute for Advanced Study, the Japan Participation Group, The Johns Hopkins University, Los Alamos National Laboratory, the Max-Planck-Institute for Astronomy (MPIA), the Max-Planck-Institute for Astrophysics (MPA), New Mexico State University, University of Pittsburgh, Princeton University, the United States Naval Observatory, and the University of Washington. % SDSS14
\end{acknowledgements}
\bibliographystyle{aa}
% \bibliography{bib_paper}

\begin{thebibliography}{35}
\expandafter\ifx\csname natexlab\endcsname\relax\def\natexlab#1{#1}\fi

\bibitem[{{Blaauw}(1961)}]{1961BAN....15..265B}
{Blaauw}, A. 1961, \bain, 15, 265

\bibitem[{{Blanton} {et~al.}(2017){Blanton}, {Bershady}, {Abolfathi},
  {Albareti}, {Allende Prieto}, {Almeida}, {Alonso-Garc{\'{\i}}a}, {Anders},
  {Anderson}, {Andrews}, \& et~al.}]{2017AJ....154...28B}
{Blanton}, M.~R., {Bershady}, M.~A., {Abolfathi}, B., {et~al.} 2017, \aj, 154,
  28

\bibitem[{{Bromley} {et~al.}(2009){Bromley}, {Kenyon}, {Brown}, \&
  {Geller}}]{2009ApJ...706..925B}
{Bromley}, B.~C., {Kenyon}, S.~J., {Brown}, W.~R., \& {Geller}, M.~J. 2009,
  \apj, 706, 925

\bibitem[{{Brown}(2015)}]{2015ARA&A..53...15B}
{Brown}, W.~R. 2015, \araa, 53, 15

\bibitem[{{Brown} {et~al.}(2015){Brown}, {Anderson}, {Gnedin}, {Bond},
  {Geller}, \& {Kenyon}}]{2015ApJ...804...49B}
{Brown}, W.~R., {Anderson}, J., {Gnedin}, O.~Y., {et~al.} 2015, \apj, 804, 49

\bibitem[{{Brown} {et~al.}(2012){Brown}, {Cohen}, {Geller}, \&
  {Kenyon}}]{2012ApJ...754L...2B}
{Brown}, W.~R., {Cohen}, J.~G., {Geller}, M.~J., \& {Kenyon}, S.~J. 2012,
  \apjl, 754, L2

\bibitem[{{Brown} {et~al.}(2014){Brown}, {Geller}, \&
  {Kenyon}}]{2014ApJ...787...89B}
{Brown}, W.~R., {Geller}, M.~J., \& {Kenyon}, S.~J. 2014, \apj, 787, 89

\bibitem[{{Brown} {et~al.}(2005){Brown}, {Geller}, {Kenyon}, \&
  {Kurtz}}]{2005ApJ...622L..33B}
{Brown}, W.~R., {Geller}, M.~J., {Kenyon}, S.~J., \& {Kurtz}, M.~J. 2005,
  \apjl, 622, L33

\bibitem[{{Brown} {et~al.}(2018){Brown}, {Lattanzi}, {Kenyon}, \&
  {Geller}}]{2018arXiv180504184B}
{Brown}, W.~R., {Lattanzi}, M.~G., {Kenyon}, S.~J., \& {Geller}, M.~J. 2018,
  ArXiv e-prints [\eprint[arXiv]{1805.04184}]

\bibitem[{{Cutri} \& {et al.}(2014)}]{2014yCat.2328....0C}
{Cutri}, R.~M. \& {et al.} 2014, VizieR Online Data Catalog

\bibitem[{{Edelmann} {et~al.}(2005){Edelmann}, {Napiwotzki}, {Heber},
  {Christlieb}, \& {Reimers}}]{2005ApJ...634L.181E}
{Edelmann}, H., {Napiwotzki}, R., {Heber}, U., {Christlieb}, N., \& {Reimers},
  D. 2005, \apjl, 634, L181

\bibitem[{{Fitzpatrick}(1999)}]{1999PASP..111...63F}
{Fitzpatrick}, E.~L. 1999, \pasp, 111, 63

\bibitem[{{Georgy} {et~al.}(2013){Georgy}, {Ekstr{\"o}m}, {Granada}, {Meynet},
  {Mowlavi}, {Eggenberger}, \& {Maeder}}]{2013A&A...553A..24G}
{Georgy}, C., {Ekstr{\"o}m}, S., {Granada}, A., {et~al.} 2013, \aap, 553, A24

\bibitem[{{Giddings}(1981)}]{1981PhDT.......113G}
{Giddings}, J.~R. 1981, PhD thesis, , University of London, (1981)

\bibitem[{{Heber} {et~al.}(2008{\natexlab{a}}){Heber}, {Edelmann},
  {Napiwotzki}, {Altmann}, \& {Scholz}}]{2008A&A...483L..21H}
{Heber}, U., {Edelmann}, H., {Napiwotzki}, R., {Altmann}, M., \& {Scholz},
  R.-D. 2008{\natexlab{a}}, \aap, 483, L21

\bibitem[{{Heber} {et~al.}(2008{\natexlab{b}}){Heber}, {Hirsch}, {Edelmann},
  {Napiwotzki}, {O'Toole}, {Brown}, \& {Altmann}}]{2008ASPC..392..167H}
{Heber}, U., {Hirsch}, H.~A., {Edelmann}, H., {et~al.} 2008{\natexlab{b}}, in
  Astronomical Society of the Pacific Conference Series, Vol. 392, Hot Subdwarf
  Stars and Related Objects, ed. U.~{Heber}, C.~S. {Jeffery}, \&
  R.~{Napiwotzki}, 167

\bibitem[{{Heber} {et~al.}(2018){Heber}, {Irrgang}, \&
  {Schaffenroth}}]{2018OAst...27...35H}
{Heber}, U., {Irrgang}, A., \& {Schaffenroth}, J. 2018, Open Astronomy, 27, 35

\bibitem[{{Hills}(1988)}]{1988Natur.331..687H}
{Hills}, J.~G. 1988, \nat, 331, 687

\bibitem[{{Hirsch} {et~al.}(2005){Hirsch}, {Heber}, {O'Toole}, \&
  {Bresolin}}]{2005A&A...444L..61H}
{Hirsch}, H.~A., {Heber}, U., {O'Toole}, S.~J., \& {Bresolin}, F. 2005, \aap,
  444, L61

\bibitem[{{Hubeny} {et~al.}(1994){Hubeny}, {Hummer}, \&
  {Lanz}}]{1994A&A...282..151H}
{Hubeny}, I., {Hummer}, D.~G., \& {Lanz}, T. 1994, \aap, 282, 151

\bibitem[{{Hummer} \& {Mihalas}(1988)}]{1988ApJ...331..794H}
{Hummer}, D.~G. \& {Mihalas}, D. 1988, \apj, 331, 794

\bibitem[{{Irrgang} {et~al.}(2014){Irrgang}, {Przybilla}, {Heber}, {B{\"o}ck},
  {Hanke}, {Nieva}, \& {Butler}}]{2014A&A...565A..63I}
{Irrgang}, A., {Przybilla}, N., {Heber}, U., {et~al.} 2014, \aap, 565, A63

\bibitem[{{Irrgang} {et~al.}(2010){Irrgang}, {Przybilla}, {Heber}, {Nieva}, \&
  {Schuh}}]{2010ApJ...711..138I}
{Irrgang}, A., {Przybilla}, N., {Heber}, U., {Nieva}, M.~F., \& {Schuh}, S.
  2010, \apj, 711, 138

\bibitem[{{Kenyon} {et~al.}(2014){Kenyon}, {Bromley}, {Brown}, \&
  {Geller}}]{2014ApJ...793..122K}
{Kenyon}, S.~J., {Bromley}, B.~C., {Brown}, W.~R., \& {Geller}, M.~J. 2014,
  \apj, 793, 122

\bibitem[{{Kurucz}(1993)}]{1993sssp.book.....K}
{Kurucz}, R.~L. 1993, {SYNTHE spectrum synthesis programs and line data}

\bibitem[{{Kurucz}(1996)}]{1996ASPC..108..160K}
{Kurucz}, R.~L. 1996, in Model Atmospheres and Spectrum Synthesis, ed. S.~J.\
  {Adelman}, F.\ {Kupka}, \& W.~W.\ {Weiss} (San Francisco: ASP), 160

\bibitem[{{Lawrence} {et~al.}(2007){Lawrence}, {Warren}, {Almaini}, {Edge},
  {Hambly}, {Jameson}, {Lucas}, {Casali}, {Adamson}, {Dye}, {Emerson},
  {Foucaud}, {Hewett}, {Hirst}, {Hodgkin}, {Irwin}, {Lodieu}, {McMahon},
  {Simpson}, {Smail}, {Mortlock}, \& {Folger}}]{2007MNRAS.379.1599L}
{Lawrence}, A., {Warren}, S.~J., {Almaini}, O., {et~al.} 2007, \mnras, 379,
  1599

\bibitem[{{L{\'o}pez-Morales} \& {Bonanos}(2008)}]{2008ApJ...685L..47L}
{L{\'o}pez-Morales}, M. \& {Bonanos}, A.~Z. 2008, \apjl, 685, L47

\bibitem[{{Marchetti} {et~al.}(2018){Marchetti}, {Contigiani}, {Rossi},
  {Albert}, {Brown}, \& {Sesana}}]{2018MNRAS.476.4697M}
{Marchetti}, T., {Contigiani}, O., {Rossi}, E.~M., {et~al.} 2018, \mnras, 476,
  4697

\bibitem[{{Poveda} {et~al.}(1967){Poveda}, {Ruiz}, \&
  {Allen}}]{1967BOTT....4...86P}
{Poveda}, A., {Ruiz}, J., \& {Allen}, C. 1967, Boletin de los Observatorios
  Tonantzintla y Tacubaya, 4, 86

\bibitem[{{Przybilla} {et~al.}(2011){Przybilla}, {Nieva}, \&
  {Butler}}]{2011JPhCS.328a2015P}
{Przybilla}, N., {Nieva}, M.-F., \& {Butler}, K. 2011, in Journal of Physics
  Conference Series, Vol. 328, Journal of Physics Conference Series, 012015

\bibitem[{{Przybilla} {et~al.}(2008{\natexlab{a}}){Przybilla}, {Nieva},
  {Heber}, {Firnstein}, {Butler}, {Napiwotzki}, \&
  {Edelmann}}]{2008A&A...480L..37P}
{Przybilla}, N., {Nieva}, M.~F., {Heber}, U., {et~al.} 2008{\natexlab{a}},
  \aap, 480, L37

\bibitem[{{Przybilla} {et~al.}(2008{\natexlab{b}}){Przybilla}, {Nieva},
  {Tillich}, {Heber}, {Butler}, \& {Brown}}]{2008A&A...488L..51P}
{Przybilla}, N., {Nieva}, M.~F., {Tillich}, A., {et~al.} 2008{\natexlab{b}},
  \aap, 488, L51

\bibitem[{{Silva} \& {Napiwotzki}(2011)}]{2011MNRAS.411.2596S}
{Silva}, M.~D.~V. \& {Napiwotzki}, R. 2011, \mnras, 411, 2596

\bibitem[{{Tremblay} \& {Bergeron}(2009)}]{2009ApJ...696.1755T}
{Tremblay}, P.-E. \& {Bergeron}, P. 2009, \apj, 696, 1755
\end{thebibliography}

%
\end{document}